\documentclass[aps,pre,twocolumn,superscriptaddress,showpacs,preprintnumbers]{revtex4-2}

\usepackage{hyperref}
\usepackage{natbib}
\usepackage{etoolbox}
\apptocmd{\thebibliography}{\raggedright}{}{} 

\usepackage[utf8]{inputenc}
\usepackage{subfiles}
\usepackage{amsfonts}
\usepackage{graphicx}
\usepackage{physics}
\usepackage{float}
\usepackage{amsmath}
\usepackage{mathtools}
\usepackage[english]{babel}
\usepackage{csquotes}

\usepackage[linecolor=red!30, disable]{todonotes}
\tikzstyle{notestyleraw} = [
	draw=black!0,
	line width=0.8pt,
	fill=red!10,
	text=black,
    text width=1.5cm,
	inner sep=0.8 ex
]

\DeclareMathOperator{\Var}{Var}

\newcommand{\veps}{\varepsilon}

\newcommand{\Kcomplexity}{$\mathcal{K}$-complexity}
\newcommand{\mCK}{\overline{C_\mathcal{K}}}

\newcommand{\siglog}{\sigma_{\text{log}}}

\newcommand{\psiz}{\ket*{\psi_0}}
\newcommand{\psit}{\ket*{\psi(t)}}

\newcommand{\psizup}{\ket*{\psi_0^{\text{up}}}}
\newcommand{\psizinf}{\ket*{\psi_0^{\text{unif}}}}
\newcommand{\psizrand}{\ket*{\psi_0^{\text{rand}}}}

\newcommand{\psizhc}{\ket*{\psi_0^{h_z=4}}}
\newcommand{\psizhz}{\ket*{\psi_0^{h_z=0}}}

\newcommand{\psizkz}{\ket*{\psi_0^{k=0}}}

\newcommand{\psizb}{\ket*{\psi_0^{\text{b}}}}

\newcommand{\mO}{\mathcal{O}}

\newcommand{\uba}{Universidad de Buenos Aires, Facultad de Ciencias Exactas y Naturales, Departamento de Física. Buenos Aires, Argentina}
\newcommand{\ifiba}{CONICET - Universidad de Buenos Aires, Instituto de Física de Buenos Aires (IFIBA). Buenos Aires, Argentina}

\begin{document}
    \title{Integrability to chaos transition through Krylov approach for state evolution}

    \author{Gast\'on F. Scialchi}
    \email[E--mail address: ]{gscialchi@df.uba.ar}
    \affiliation{\uba}
    \affiliation{\ifiba}

    \author{Augusto J. Roncaglia}
    \affiliation{\uba}
    \affiliation{\ifiba}

    \author{Diego A. Wisniacki}
    \email[E--mail address: ]{wisniacki@df.uba.ar}
    \affiliation{\uba}
    \affiliation{\ifiba}

    \begin{abstract}
        The complexity of quantum evolutions can be understood by examining their dispersion in a chosen basis. Recent research has stressed the fact that the Krylov basis is particularly adept at minimizing this dispersion [V. Balasubramanian et al,
    Physical Review D 106, 046007 (2022)]. This property assigns a central role to the Krylov basis in the investigation of quantum chaos. Here, we delve into the transition from integrability to chaos using the Krylov approach, employing an Ising spin chain and a banded random matrix model as our testing models. Our findings indicate that both the saturation of Krylov complexity and the dispersion of the Lanczos coefficients can exhibit a significant dependence on the initial condition. However, both quantities can gauge dynamical quantum chaos with a proper choice of the initial state.
    \end{abstract}

    \maketitle

    \section{Introduction}
    \label{sec:introduction}
The field of quantum chaos has made significant advancements in unraveling the intricacies of quantum systems by characterizing their spectral properties through the utilization of random matrix models associated with various statistical ensembles \cite{mehta2004random, wimberger2014nonlinear}.  This statistical framework has also proven invaluable in portraying the attributes of quantum systems with many-body interactions, where a classical depiction is elusive \cite{richter2022semiclassical}. Consequently, a measure of chaos in these systems typically revolves around their spectral patterns.
In conjunction with this spectral perspective on quantum chaos, an additional venue of exploration delves into the analysis of their dynamics. This entails exploring measures like the Loschmidt echo and out-of-time-ordered correlators (OTOCs) which provide insight into the system's irreversibility when subjected to perturbations and the propagation behavior of operators as they evolve in time \cite{gorin2006dynamics,Wisniacki2012,swingle2018unscrambling,Garcia-Mata:2023}.

A novel addition to the dynamic approach is known as Krylov complexity (\Kcomplexity) \cite{Parker2019}. This measure was developed for both Schrödinger evolution of states \cite{Balasubramanian_2022} and Heisenberg evolution of observables in an operator Hilbert space formed via a suitable inner product \cite{Rabinovici2021}. It is derived using the Lanczos algorithm that, given an initial condition, yields a series of values referred to as Lanczos coefficients and an orthonormal basis for the Krylov subspace that encompasses the system evolution. This algorithm transforms the original system into a one-dimensional tight-binding problem with disordered attributes, where the hopping amplitudes are determined by the Lanczos coefficients. The concept of \Kcomplexity\ is then established as the dispersion of the one-particle wavefunction across the Krylov space.

Under operator formalism,
it has been hypothesized \cite{Parker2019} that generic chaotic systems in the thermodynamic limit exhibit an exponential growth of \Kcomplexity, which sharply bounds the Lyapunov exponent as defined through the growth of OTOCs.
Further research has been done in finite systems \cite{barbon2019evolution, Rabinovici2021, Rabinovici_2022_localization, Rabinovici_2022_integrability},
where it was conjectured and observed that chaotic systems exhibit a larger long-time saturation value of \Kcomplexity\ when compared to integrable ones,
and that this can be understood as Anderson localization in the associated tight-binding problem
due to the Lanczos sequences derived from integrable systems having more disorder.
However, this seems not to be universal due to its strong dependency on
the chosen initial operator \cite{PhysRevE.107.024217}.

The goal of this study is to explore the transition from integrability to chaos using the Krylov formalism applied to quantum states \cite{Balasubramanian_2022}. To achieve this goal, we examine the dynamics of \Kcomplexity\ saturation, Lanczos coefficient dispersion, and how these vary in relation to the level of chaos present within the system.
We do this by performing a numerical study of the aforementioned quantities along with the parametric integrability-to-chaos transition present in both an Ising spin chain subject to an external magnetic field
and a banded random model \cite{banuls2011strong,li2017measuring,mirkin2021quantum,chavda2014poisson}.
Our research shows that both the saturation of Krylov complexity and the dispersion of Lanczos coefficients are influenced by the initial conditions. However,
they can effectively assess the onset of dynamical quantum chaos when the initial state is carefully selected.
We also found that the dispersion of the Lanczos coefficients shows a more robust behaviour for the initial conditions considered in this work.

This work is organized as follows.
In Sec. \ref{sec:lanczos_and_complexity} we briefly go over the Lanczos algorithm, define the Krylov complexity, and review both the relation between its long-time saturation value  and the dispersion of Lanczos coefficients with chaos.
In Sec. \ref{sec:ising} we present our numerical results for an Ising spin chain with a longitudinal-transverse magnetic field
using a variety of initial states
and explore the best choice in order to extract universal features of the system's complexity.
An analogous study for a random banded model is shown in Sec. \ref{sec:random},
in order to test how these results behave in the integrability-to-chaos transition for a generic system modeled through its spectral statistics.
To highlight the importance of the choice of the initial state in the Krylov approach,  in App. \ref{sec:saturation_state_dependence} we analytically study the dependence of the Krylov complexity saturation with the
energy localization of the initial state.
We conclude in Sec. \ref{sec:discussion} with a summary of our results and some final remarks.

\section{Lanczos algorithm and Krylov complexity saturation}
\label{sec:lanczos_and_complexity}

As an initial state $\psiz$ evolves in time, it traverses a fraction of the system Hilbert space known as the Krylov subspace. When considering the evolution under a time-independent Hamiltonian $H$:
\begin{equation}
    \psit = e^{-it H} \psiz = \sum_{n=0}^\infty \frac{(-it)^n}{n!} H^n \psiz.
    \label{eq:psit_formal_expansion}
\end{equation}
From this expression it is possible to obtain a basis for this subspace. For systems with a finite Hilbert space dimension $D$, only a finite number (denoted as $K$) of the states $H^n \psiz$ can be linearly independent. Therefore, the Krylov subspace is the space spanned by the successive applications of the Hamiltonian to the initial state,
\begin{equation}
    \mathcal{K} = \text{span}\{H^n\psiz\}_{n=0}^{K-1}.
    \label{eq:k_subspace_span}
\end{equation}
An orthonormal basis for $\mathcal{K}$ can be obtained  employing the Lanczos algorithm \cite{parlett1998symmetric, viswanath2008recursion} which recursively applies the Gram-Schmidt procedure to Eq.~\eqref{eq:k_subspace_span}.
In this way, the Krylov basis elements $\{\ket*{K_n}\}_{n=0}^{K-1}$ are defined through
\begin{equation}
    \ket*{A_{n+1}} = (H - a_n)\ket*{K_n} - b_n\ket*{K_{n-1}}, \quad \ket*{K_n} = b_n^{-1} \ket*{A_n},
    \label{eq:lanczos_alg}
\end{equation}
where $a_n$ and $b_n$ are the Lanczos coefficients:
\begin{equation}
    a_n = \bra*{K_n} H \ket*{K_n}, \quad b_n = \sqrt{\braket{A_n}{A_n}},
    \label{eq:lanczos_coefs}
\end{equation}
obtained in terms of the initial conditions $b_0 = 0$ and $\ket*{K_0} = \psiz$.
This algorithm also transforms the Hamiltonian into a tridiagonal form,
\begin{equation}
    H \ket{K_n} = a_n \ket{K_n} + b_{n+1} \ket{K_{n+1}} + b_{n} \ket{K_{n-1}},
    \label{eq:h_tridiagonal}
\end{equation}
where $a_n$ and $b_n$ represent the diagonal and off-diagonal elements of the Hamiltonian in the Krylov basis.
The Lanczos algorithm, as delineated in Eq.~\eqref{eq:lanczos_alg}, can exhibit numerical instability due to the accumulation of roundoff errors stemming from floating-point computations. Various alternative implementations have been proposed to mitigate this issue, as outlined in \cite{parlett1998symmetric, simon1984lanczos}. Here, we adopt the full orthogonalization variant of the algorithm.

The time-evolved state [Eq.~\eqref{eq:psit_formal_expansion}] expanded within the Krylov basis is:
\begin{equation}
    \ket{\psi(t)} = \sum_{n=0}^{K-1} \psi_n(t) \ket{K_n},
    \label{eq:psit_krylov_basis}
\end{equation}
Then, incorporating Eq.~\eqref{eq:h_tridiagonal} into the Schrödinger equation, an expression governing the evolution of the coefficients $\psi_n(t)$ can be derived,
\begin{equation}
    i\partial_{t} \psi_n(t) = a_n \psi_n(t) + b_{n}\psi_{n-1}(t) +b_{n+1}\psi_{n+1}(t).
    \label{eq:psit_tightbinding}
\end{equation}
This expression indicates that the evolution of $\psiz$ can be obtained in terms of that of a one-dimensional tight-binding problem.
With this analogy, the hopping and site coefficients are dictated by the Lanczos sequences,
and the initial condition of the tight-binding problem is completely localized at the leftmost site of the chain ($\psi_n(0) = \delta_{n0}$).
Starting from $t=0$, where only the element $\ket*{K_0}$ of the Krylov basis is relevant,
the wavefunction $\psi_n(t)$ progressively permeates further into the expanse of the Krylov space as time goes on.
This expansion prompts the requirement for an increasing number of basis elements to faithfully capture the unfolding evolution.
This realization naturally leads to the concept of evolution complexity referred to as Krylov complexity
\cite{Parker2019, Rabinovici2021},
which is defined as the average spatial position of the wavefunction throughout the chain,
\begin{equation}
    C_{\mathcal{K}}(t) = \sum_{n=0}^{K-1} n \abs{\psi_n(t)}^2.
    \label{eq:k_complexity}
\end{equation}
This value encapsulates the average dimension of the Krylov subspace representing the evolution from the initial state up to time $t$.
We note that, with the exception of exact degeneracies in the Hamiltonian spectrum caused by symmetries, the Krylov subspace that encapsulates the evolution of the system for all times covers the entire Hilbert space in all the scenarios that we have studied. Furthermore, we have observed that the presence of such exact degeneracies can introduce numerical instability in the Lanczos algorithm, even when employing the full orthogonalization routine. This challenge can be circumvented by confining the Hamiltonian to a symmetry subspace.
We further assume an absence of accidental degeneracies in the spectrum of the Hamiltonian.
For this reason, we set $K = D$, understanding that the Krylov basis expands the full Hilbert space or a symmetry subspace of dimension $D$.

In this work we are mainly interested on the longtime saturation value of the \Kcomplexity~which, by a procedure completely analogous to the one presented in Ref.~\cite{Rabinovici_2022_localization}, can be shown to follow
\begin{equation}
    \mCK \equiv \lim_{T\to\infty} \frac{1}{T} \int_{0}^{T}dt\, C_{\mathcal{K}}(t)
        = \sum_{n=0}^{D-1} n Q_{0n},
    \label{eq:k_complexity_saturation}
\end{equation}
where
\begin{equation}
    Q_{0n} \equiv \sum_{i=1}^D \abs{\braket{e_i}{\psi_0}}^2 \abs{\braket{K_n}{e_i}}^2
    \label{eq:q0n}
\end{equation}
and $H \ket*{e_i} = e_i \ket*{e_i}$.
By expanding the eigenstates of the Hamiltonian in the Krylov basis
$\ket{e_i} = \sum_{n=0}^{D-1} \veps^i_n \ket*{K_n}$
and using Eq.~\eqref{eq:h_tridiagonal} we obtain a related tight-binding problem:
\begin{equation}
    e_i \veps^i_n = a_n \veps^i_n + b_{n}\veps^i_{n-1} +b_{n+1}\veps^i_{n+1},
    \label{eq:h_evec_tightbinding}
\end{equation}
 where  $\abs{\veps^i_n}^2 = \abs{\braket{K_n}{e_i}}^2$.

Under the operator formalism (where $a_n = 0$ $\forall n$)
it was proposed in \cite{Rabinovici_2022_localization}, and observed in certain systems in \cite{Rabinovici_2022_integrability},
that the saturation value [Eq.~\eqref{eq:k_complexity_saturation}] should be larger for chaotic systems than for integrable ones.
It is argued heuristically that an abundance of quasidegeneracies in the spectrum of the system leads to an erratic behavior of the Lanczos coefficients $b_n$.
As a consequence, the relation in Eq.~\eqref{eq:h_evec_tightbinding} describes a tight-binding problem with off-diagonal disorder,
which is subject to Anderson localization.
As integrable systems are typically characterized by Poissonian level spacing statistics,
they present a high degree of quasidegeneracies leading to a
larger disorder.
Conversely, the level repulsion present in chaotic systems implies
less disorder,
and a more delocalized wavefunction.
The same argument may be extended to the state formalism
with the addition of diagonal disorder.
We then have, in the completely delocalized limit, $\abs{\veps^i_n}^2 = 1/D$ $\forall i, n$
and from Eq.~\eqref{eq:k_complexity_saturation}:
\begin{equation}
    \mCK = \frac{1}{D} \sum_{n=0}^{D-1} n = \frac{D -1}{2}.
    \label{eq:k_complexity_saturation_delocalized}
\end{equation}

\begin{figure}[ht]
    \centering
    \includegraphics[width=\linewidth]{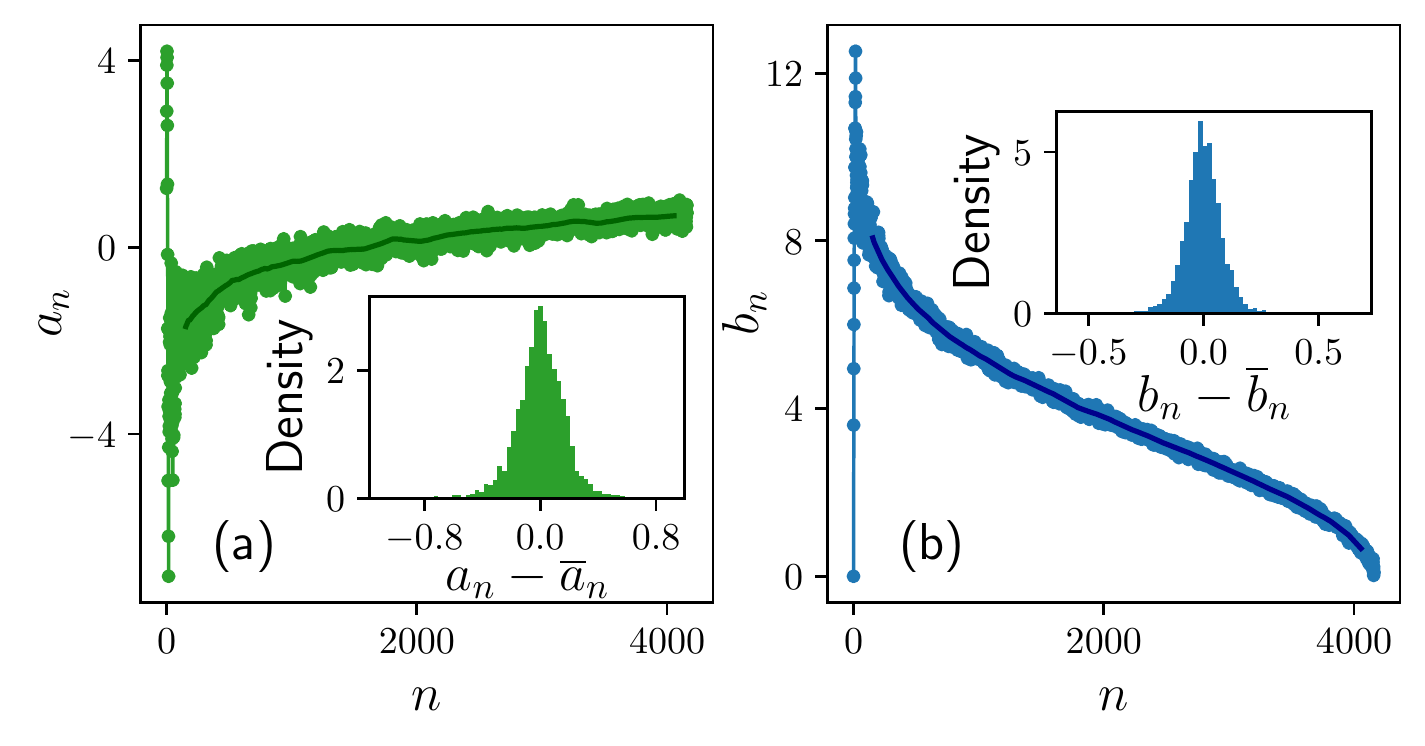}
    \caption{Example of the (a) diagonal and (b) off-diagonal Lanczos sequences and their moving averages (dark lines).
    The insets show the distribution of the fluctuations of the sequences about said moving averages.
    These sequences correspond to the Ising model discussed in Sec.~\ref{sec:ising}
    with a value of the magnetic field component $h_z = 1.02$.
}
    \label{fig:coefs_dispersion_example}
\end{figure}

\begin{figure}[ht]
    \centering
    \includegraphics[width=\linewidth]{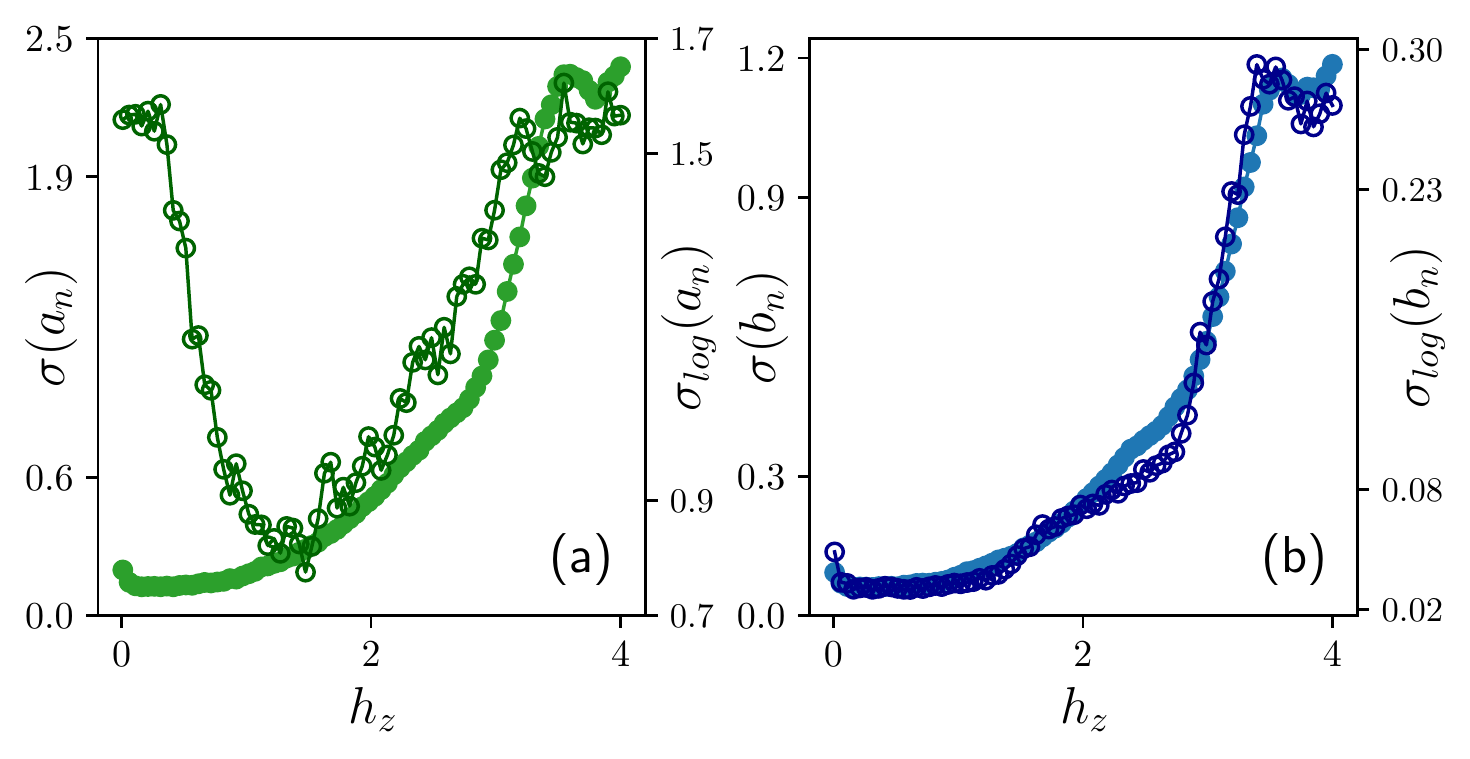}
    \caption{Comparison of Lanczos coefficient dispersion measures
        applied to the (a) diagonal and (b) off-diagonal sequences
        as a function of the magnetic field component $h_z$
        in the Ising model discussed in Sec.~\ref{sec:ising}.
        The filled circles correspond to the dispersion definition in Eq. \eqref{eq:disp_mov},
        while the dark empty circles to that of Eq. \eqref{eq:loc_disp_log}.
    }
    \label{fig:coefs_dispersion_comparison}
\end{figure}

In this study, we also aim to investigate the relation between chaotic behavior and the dispersion of the Lanczos coefficients under state formalism.
The localization length $L_{\text{loc}}$ of the wavefunction has been proposed to be related to the amount of disorder in the off-diagonal sequence through~\cite{fleishman1977fluctuations}
\begin{equation}
    L_{\text{loc}} \propto \siglog^{-1}(b_n),
    \ \text{where} \
    \siglog^2(b_n) = \Var\left( \ln \left|\frac{b_{2n-1}}{b_{2n}}\right| \right)
    \label{eq:loc_disp_log}.
\end{equation}
This dispersion measure can suffer from large fluctuations when the Lanczos sequence is small,
which may be the case for $a_n$ since it typically fluctuates around zero for sufficiently large $n$
(as Fig.~\ref{fig:coefs_dispersion_example}.(a) shows).
An intuitive alternative definition is to measure the dispersion of the sequences with respect to their local mean.
Given a sequence $s_n$ of length $N$, the dispersion about its moving average is
\begin{equation}
    \sigma^2(s_n) \equiv \frac{1}{N} \sum_{n=n_0}^N (s_n - \overline{s}_n)^2,
    \ \text{with} \
    \overline{s}_n = \frac{1}{2w} \sum_{m=n-w}^{n+w} s_m,
    \label{eq:disp_mov}
\end{equation}
where $w$ is the half-width of the window used to compute the moving average
and $n_0$ the value of the sequence index from which to start computing the dispersion.
We show in Fig.~ \ref{fig:coefs_dispersion_example}
an example of the quantities just described
applied to an instance of Lanczos sequences.
For the data in the figure
(and for the rest of the results in this work)
we chose $w = 0.025 N$,
and a value of $n_0$ so as to avoid the initial ramp in the sequences.
A comparison between both measures of dispersion can be seen in Fig.~\ref{fig:coefs_dispersion_comparison}.
For the sequence $b_n$, both definitions \eqref{eq:loc_disp_log} and \eqref{eq:disp_mov} agree
[see Fig.~\ref{fig:coefs_dispersion_comparison}.(b)],
but it is not the case when applied to $a_n$ [Fig.~\ref{fig:coefs_dispersion_comparison}.(a)].
Moreover, the functional shape of $\sigma(a_n)$ is almost identical to that of $\sigma(b_n)$,
except for a rescaling factor.
The data shown in  Figs.~\ref{fig:coefs_dispersion_example} and \ref{fig:coefs_dispersion_comparison}
were obtained using the Ising model discussed in Sec.~\ref{sec:ising}
restricted to the positive parity subspace,
and the initial state with all $N=13$ spins in the 'up' configuration.

\section{Gauging quantum complexity through Krylov evolution of states}
\label{sec:num_results}

Here we present numerical results for the quantities detailed in the previous section.
Namely, the saturation of Krylov complexity [Eq.~\eqref{eq:k_complexity_saturation}] and
the dispersion of the Lanczos coefficients [Eq.~\eqref{eq:disp_mov}].
We trace these quantities along
the integrability to chaos transition of an Ising spin chain in Sec.~\ref{sec:ising}
and
an statistical ensemble transition for a banded random matrix model in Sec.~\ref{sec:random},
comparing them with the spectral statistics of each system as a quantifier of quantum chaos.
This is done for various initial states.

The spectral measure of chaos that we used
is the mean value of the ratio of consecutive energy level spacings $s_i=e_i-e_{i-1}$ \cite{atas2013distribution},
\begin{equation}
    \langle \tilde r \rangle = \frac{1}{D} \sum_{n=1}^D \tilde r_n,
    \quad \text{where} \quad
    \tilde r_n = \frac{\min(s_n, s_{n-1})}{\max(s_n, s_{n-1})}
    \label{eq:r_tilde},
\end{equation}
which takes the values
$\langle \tilde r \rangle_{P} \approx 0.38629$
for Poissonnian statistics and
$\langle \tilde r \rangle_{WD} \approx 0.53590$ for the orthogonal Gaussian ensemble.
We normalize this quantity as
\begin{equation}
    \eta = \frac{\langle \tilde r \rangle - \langle \tilde r \rangle_{P}}{\langle \tilde r \rangle_{WD} - \langle \tilde r \rangle_{P}}
    \label{eq:chaometer}
\end{equation}
to obtain a measure of quantum chaos such that
$\eta \approx 1$ for a chaotic system and $\eta \approx 0$ for an integrable system.

\subsection{Integrability-to-chaos transition in a spin chain}
\label{sec:ising}

We consider an Ising model with nearest-neighbor interactions in the $z$ direction and a magnetic field on the $x-z$ plane given by the Hamiltonian
\begin{equation}
H = \sum_{i=1}^N (\sigma_x^{(i)} + h_z \sigma_z^{(i)})
        - \sum_{i=1}^{N-1} \sigma_z^{(i)}\sigma_z^{(i+1)}
    \label{eq:h_ising},
\end{equation}
where $N$ is the number of spin-1/2 particles in the chain
and $\sigma_k^{(i)}$ the Pauli operator at the $i$-th site in the direction $k$.
This model exhibits an integrability-to-chaos  transition as a function of the longitudinal component of the magnetic field $h_z$, such that the system is integrable for $h_z = 0$ and $h_z \gg 1$,
and chaotic when $h_z \sim 1$.

We also consider open boundary conditions,
which makes the system invariant under reflection with respect to the center of the chain.
This implies that the Hamiltonian commutes with the parity operator,
allowing it to be decomposed into disconnexe even and odd parity subspaces
of dimension $D^{\text{even/odd}} \simeq D/2$
such that the dimension of the full Hilbert space is $D = D^{\text{even}} + D^{\text{odd}}$.
Here we restrict the dynamics to the even subspace.
For $h_z = 0$ the system acquires an additional symmetry
associated to the conservation of the magnetization,
but we avoid this point in our calculations.

\begin{figure}[ht]
    \centering
    \includegraphics[width=\linewidth]{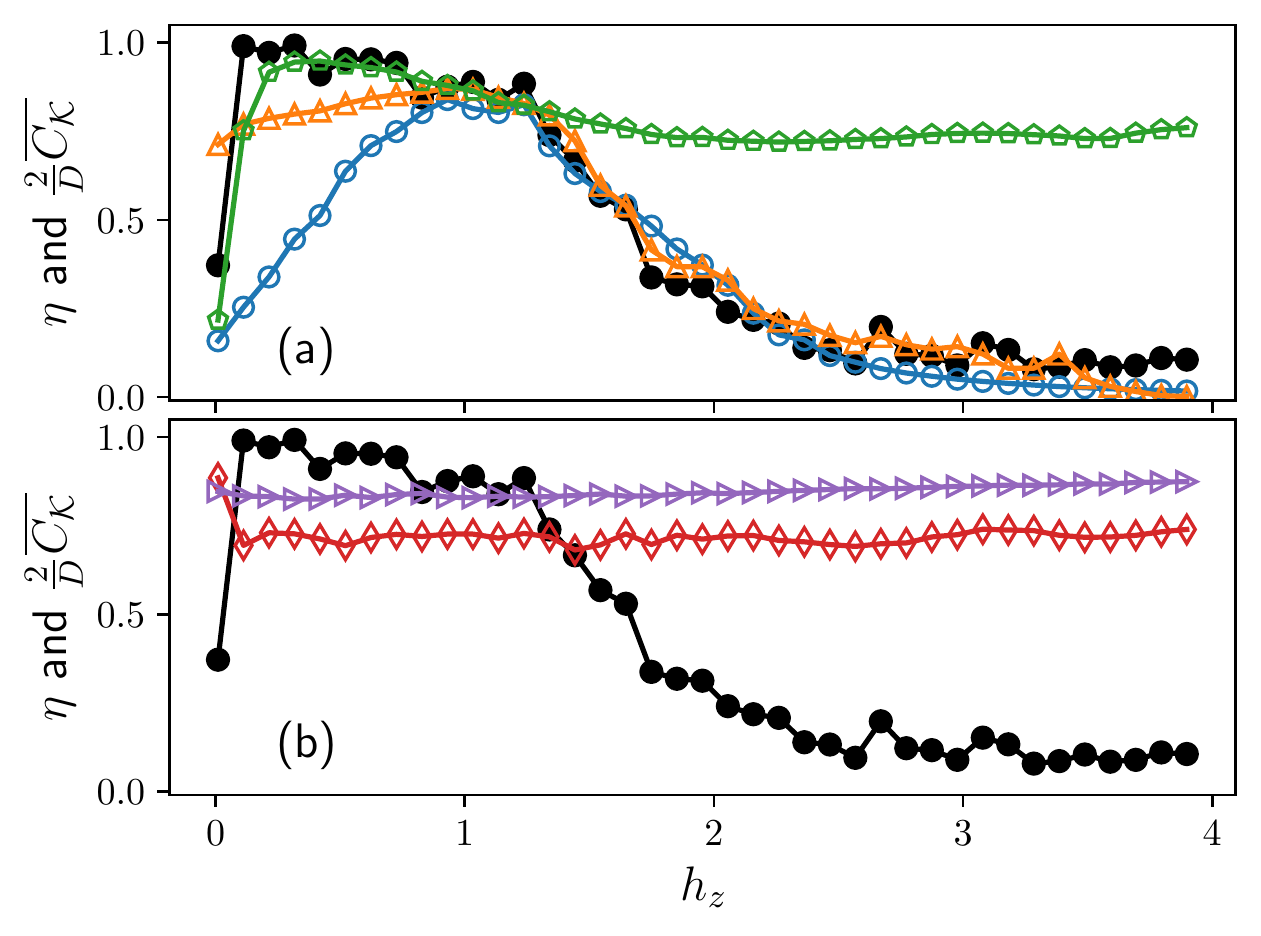}
    \caption{Chaotic measure $\eta$ (black filled circles) and saturation of the \Kcomplexity\
        as a function of the chaos parameter $h_z$ through the chaos-to-integrability transition in the Ising spin chain for a variety of initial states.
        The complexity curves  correspond to
        (a)
        an initial state in the all spins 'up' configuration (blue circles),
        an average over initial states from the eigenbasis of the integrable Hamiltonian with $h_z = 4$ (orange triangles), and an
        average over initial states from the eigenbasis of the integrable Hamiltonian with $h_z = 0$ (green pentagons),
        (b)
        average over random initial states (red diamonds)
        and
        an initial state uniformly distributed over the eigenbasis of the integrable Hamiltonian with $h_z = 0$ (purple triangles).
    }
    \label{fig:Ising_saturations}
\end{figure}

\begin{figure}[ht]
    \centering
    \includegraphics[width=\linewidth]{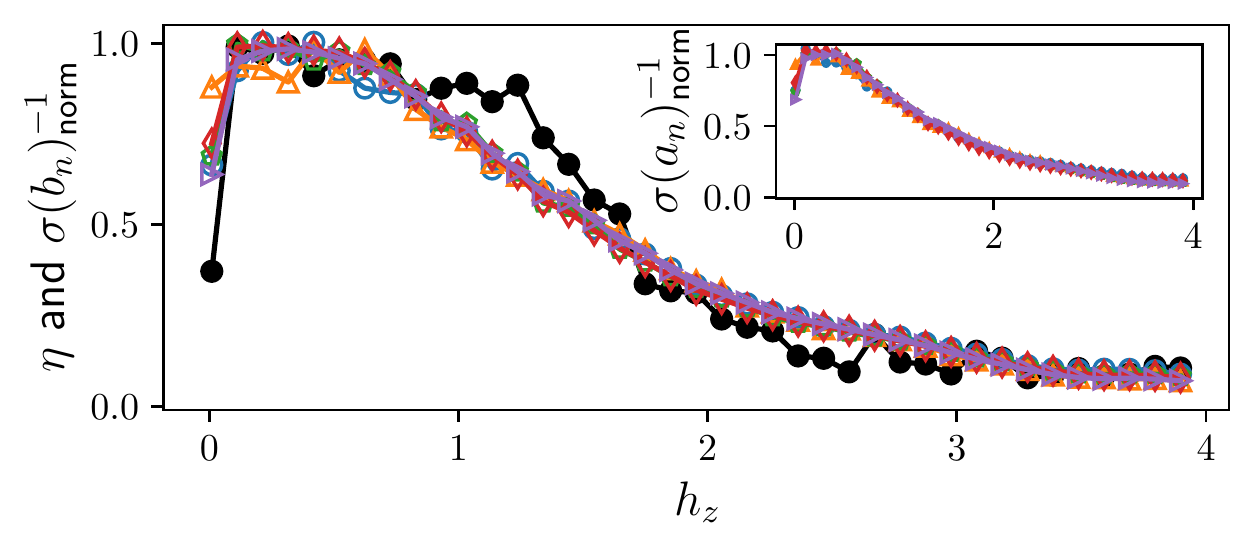}
    \caption{Chaotic measure $\eta$ (black filled circles) and normalized inverse of the
        dispersion of Lanczos coefficients as a function of the chaos parameter $h_z$
        through the chaos-to-integrability transition in the Ising spin chain for a variety of initial states.
        The main plot shows the dispersion of the off-diagonal sequence $b_n$,
        while the inset that of the diagonal $a_n$
        (which has almost identical curves).
      Symbols of each curve are the same as in Fig.~\ref{fig:Ising_saturations}.
        }
    \label{fig:Ising_dispersions}
\end{figure}

We have performed simulations for different values of $h_z$ and considered different initial states.
Figure~\ref{fig:Ising_saturations} shows
the saturation of \Kcomplexity\ [Eq.~\eqref{eq:k_complexity_saturation}]
normalized by its approximate expected value for full delocalization in Krylov space
[Eq.~\eqref{eq:k_complexity_saturation_delocalized}].
In Figure~\ref{fig:Ising_dispersions}
it is shown the inverse dispersion of the Lanczos coefficients [Eq.~\eqref{eq:disp_mov}] with
a normalization that is detailed in Appendix~\ref{sec:normalization}.
Both quantities are compared with the chaoticity of the system as defined through Eq.~\eqref{eq:chaometer}.
The curve for $\eta$ was obtained by setting $N=13$,
while the complexity saturation and dispersion correspond to $N=10$.
The latter two remain invariant upon modifying the number of spins.
The initial states we considered are:
all spins in the 'up' state $\psizup$,
a state uniformly spread over the energy eigenbasis with $h_z = 0$ $\psizinf$,
random states $\{\psizrand\}$,
eigenstates of the integrable Hamiltonian with $h_z=4$ $\{\psizhc\}$, and
eigenstates of the integrable Hamiltonian with $h_z=0$ $\{\psizhz\}$.
In the last three cases
we considered averages over the respective sets of states.
For the random states the average was made over ten realizations of the initial state,
and in the remaining two cases over forty states around the center of the energy spectrum
of each respective integrable Hamiltonian.
The results remain
largely invariant as we increment the number of states in these averages.

As can be seen in Fig.~\ref{fig:Ising_saturations},
the saturation value of Krylov complexity exhibits a high dependence on the initial state.
The curves corresponding to the states $\psizup$ and the average over $\{\psizhc\}$ [Fig.~\ref{fig:Ising_saturations}.(a)]
display a remarkable correlation with the chaoticity quantifier $\eta$,
although the former exhibits a decrease in complexity as $h_z$ goes to zero
even in regions where the system is still highly chaotic.
For the average over the set $\{\psizhz\}$ [Fig.~\ref{fig:Ising_saturations}.(a)],
the saturation value of the Krylov complexity is high
except in the region $h_z \approx 0$.
On the other hand, when examining the uniformly distributed state and the average over random states [Fig.~\ref{fig:Ising_saturations}.(b)],
there is no evident relationship between chaoticity and Krylov complexity saturation. In Appendix \ref{sec:saturation_state_dependence}, we analytically discuss the Krylov complexity saturation for uniformly distributed and energy-localized
initial states.

The dependency of the saturation of \Kcomplexity\ on the initial state contrasts with the behavior of the dispersion of the Lanczos coefficients, which varies less with the initial conditions studied in this work, as Fig.~\ref{fig:Ising_dispersions} shows.
This observation applies for both the diagonal $a_n$ and off-diagonal sequence $b_n$,
and it is interesting to note that both also share a similar functional relation with respect to $h_z$,
which holds true for all cases we consider in this study.
More importantly, we observe that the inverse of the dispersions of both diagonal $a_n$ and off-diagonal $b_n$ elements exhibit a striking resemblance to the spectral chaos measure $\eta$. In essence, they provide a more robust and accurate description of the chaos-integrability transition, surpassing the Krylov complexity, which is much more sensitive to the initial state.

Notice that the relation between the dispersion of the Lanczos coefficients and $\eta$ that we have shown in this work is different from the one that  was found in Ref.~\cite{PhysRevE.107.024217} under the Heisenberg picture.
There, the chaotic transition as measured by $\eta$ could be traced with $-\sigma$,
while our current findings under the Schrödinger picture indicate that an accurate description  can be achieved through the inverse of the dispersion $1/\sigma$.
It would be interesting to explore  the conection between the Krylov formalism within the Heisenberg picture and its counterpart within the Schrödinger picture.

For $h_z=0$ the system gains an additional symmetry,
and as such we have avoided this particular point.
Nevertheless, its effects are still reflected on the spectral statistics in a small $h_z$ region around the symmetry point.
It is interesting to note that this feature also seems to have an impact on the saturation of \Kcomplexity\ [Fig.~\ref{fig:Ising_saturations}.(a)],
and on the dispersion of the Lanczos coefficients
[Fig.~\ref{fig:Ising_dispersions}].

While it is not shown in this work, we have also observed that
when the initial state is an eigenstate of the operator
$\sum_{i=1}^{N-1} \sigma_{k}^{(i)}\sigma_{k}^{(i+1)}$ with $k = x, y$,
both the \Kcomplexity\ saturation and the dispersion of Lanczos coefficients exhibit similarities to those observed  in Figs.~\ref{fig:Ising_saturations}.(b) and \ref{fig:Ising_dispersions}, corresponding to $\psizinf$ and $\{\psizrand\}$ respectively.
This suggests that it is not required to select a fully delocalized or random state in order for the complexity saturation to fail in  detecting the chaotic transition.

\subsection{Integrability-to-chaos transition in a banded random model}
\label{sec:random}

To investigate the behavior of these quantities in generic systems, we consider a random matrix model.
In particular, we study Hamiltonians obtained from banded random matrices of the form
\begin{equation}
    H_{\text{RMT}} = \frac{H_0 + k V}{\sqrt{1 + k^2}}
    \label{eq:h_random_banded},
\end{equation}
where $H_0$ is a diagonal matrix with elements drawn from a normal distribution with zero mean and unity variance, and $V$ a real banded random matrix of bandwidth $b$ \cite{chavda2014poisson}.
Hence, this model exhibits a transition from Poissonian level statistics at $k=0$ to GOE when $k$ reaches a sufficiently large value. This is associated to the transition from a system with  integrable dynamics to one that exhibits chaotic behavior.
The dependence of $H_{\text{RMT}}$ with $k$ ensures that its  spectrum remains bounded~\cite{PhysRevLett.65.2325, PhysRevA.44.8043}.

\begin{figure}[ht]
    \centering
    \includegraphics[width=\linewidth]{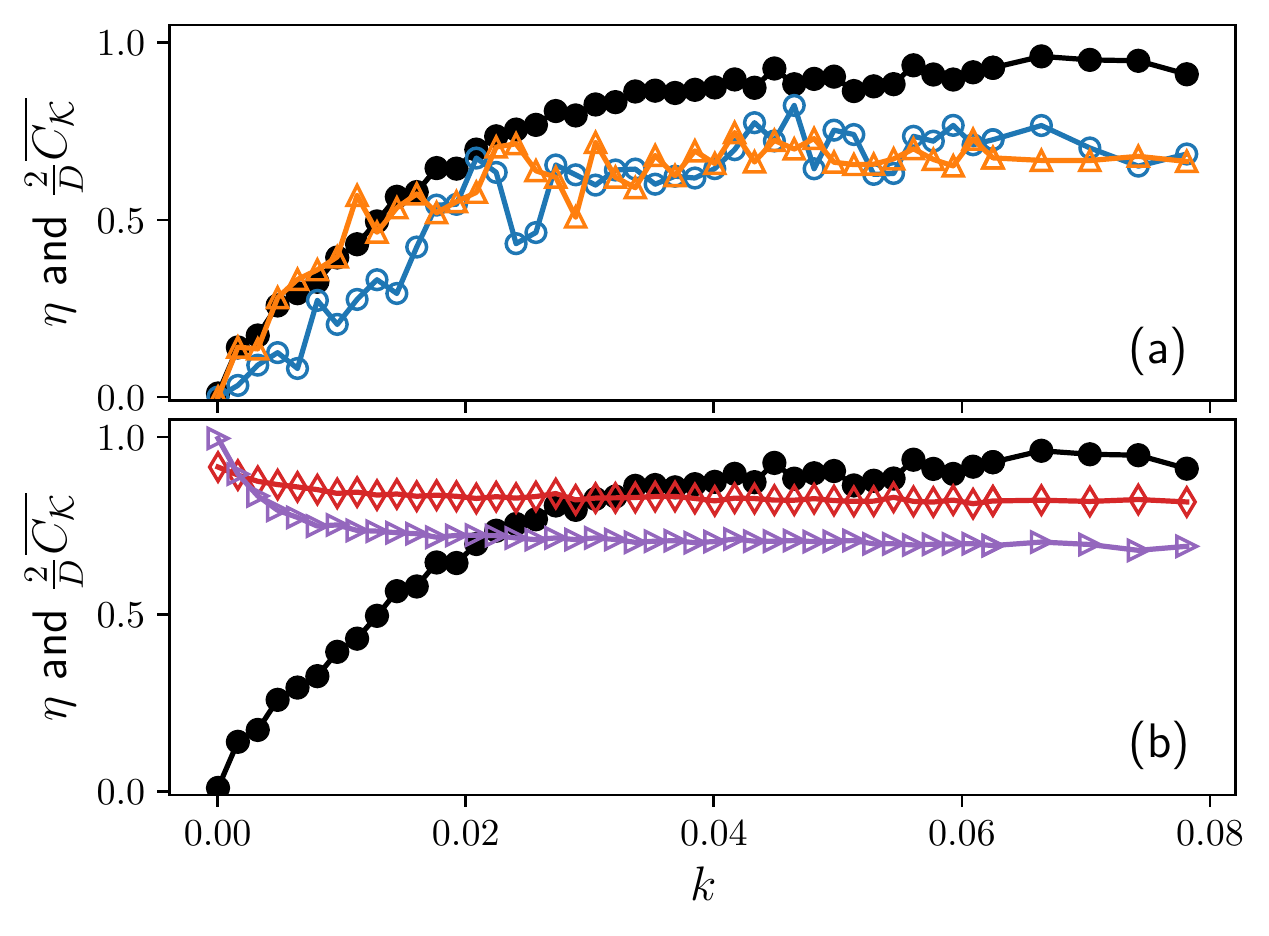}
    \caption{Chaotic measure $\eta$ (black filled circles) and saturation of the \Kcomplexity\
        as a function of the chaos parameter $k$ through the Poisson-to-GOE ensemble transition
        for the banded random matrix model
        for a variety of initial states.
        The curves for the Krylov complexity correspond to
        (a)
        an initial state that is an eigenstate of $H_{\text{RMT}}$ with $k=0$ in the border of its spectrum (blue circles),
        an average over initial states from the eigenbasis of $H_{\text{RMT}}$ with $k=0$ (orange triangles),
        (b)
        an average over random initial states (red diamonds)
        and
        an initial state uniformly distributed over the eigenbasis of $H_{\text{RMT}}$ with $k=0$ (purple triangles).
    }
    \label{fig:random_saturations}
\end{figure}

\begin{figure}[ht]
    \centering
    \includegraphics[width=\linewidth]{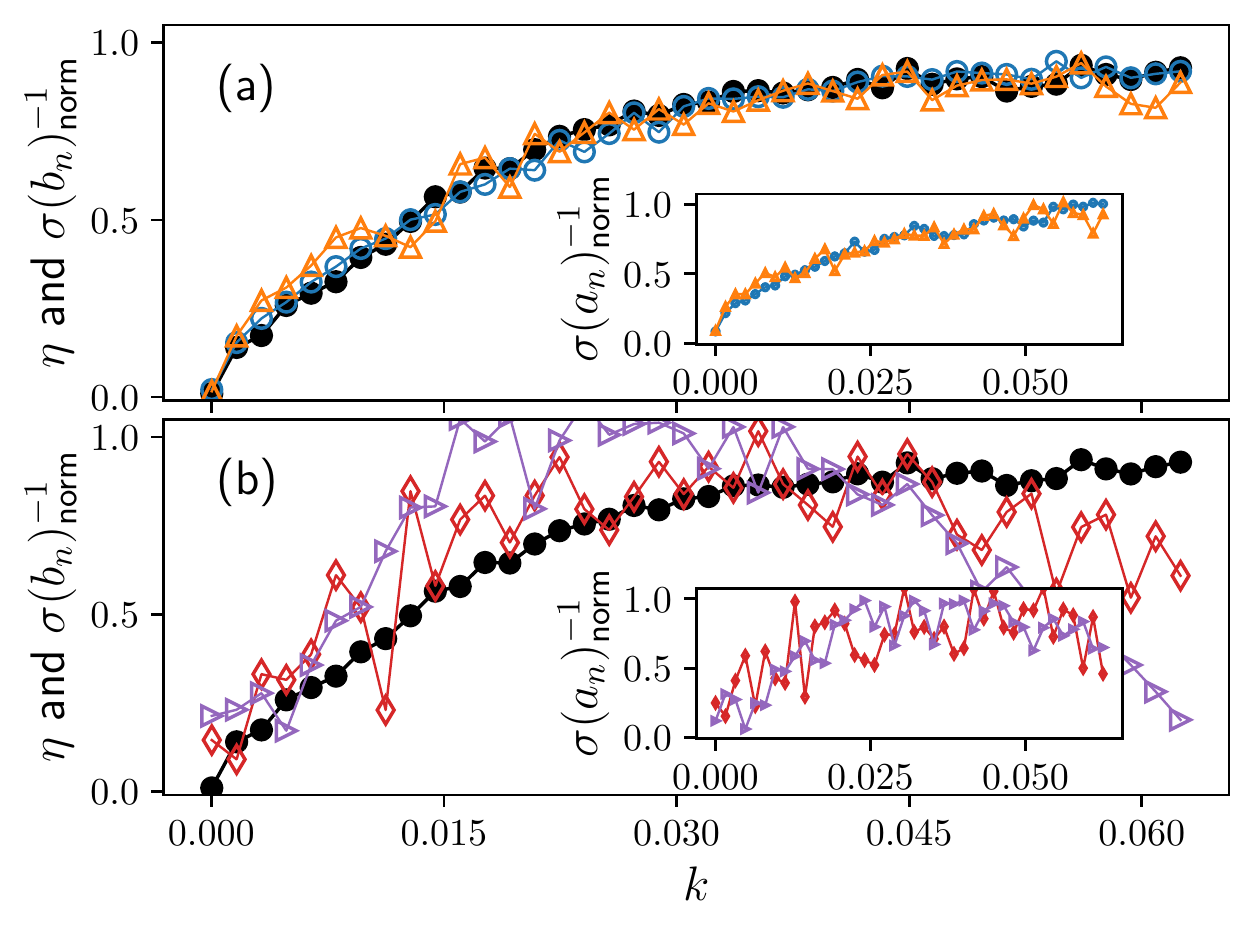}
    \caption{Chaotic measure $\eta$ (black filled circles) and normalized inverse of the
        dispersion of Lanczos coefficients
        as a function of the chaos parameter $k$ through the Poisson-to-GOE ensemble transition
        for the banded random matrix model
        for a variety of initial states.
        The main plots show the dispersion of the off-diagonal sequence $b_n$,
        while the insets that of the diagonal $a_n$
        (which has almost identical curves).
        Symbols for each curve and panel distribution are the same as in Fig.~\ref{fig:random_saturations}.
}
    \label{fig:random_dispersions}
\end{figure}

The results for this model are shown in Figs.~\ref{fig:random_saturations} and \ref{fig:random_dispersions},
and were obtained for a Hilbert space dimension of $D = 1024$ with a bandwidth $b = 0.2 D$.
Similarly to the previous approach, we have chosen the following initial states:
an eigenstate of the Hamiltonian with $k=0$ in the border of its spectrum $\psizb$,
the uniformly spread state in the eigenbasis of the Hamiltonian with $k=0$ $\psizinf$,
random states $\{\psizrand\}$
and
eigenstates of the Hamiltonian with $k=0$ $\{\psizkz\}$.
For the first two,
we considered
an average over ten realizations of the random Hamiltonian.
For the last two cases
only one realization of the Hamiltonian was considered, while
 the averages were taken
over ten random states,
and twenty states around the center of the spectrum,
respectively.

In Fig.~\ref{fig:random_saturations}, we can observe a relation of the \Kcomplexity\ with the initial state that is similar to what was found for the Ising model.
While some states exhibit a discernible transition, others do not show any clear correlation.
Namely, the \Kcomplexity\ corresponding to the states $\psizb$ and  $\{\psizkz\}$ is able to detect the transition [Fig~\ref{fig:random_saturations}.(a)].
On the other hand,  for both the uniformly distributed state $\psizinf$ and the average over random states $\{\psizrand\}$ we observe a relatively high value of \Kcomplexity\ that is mostly insensitive to the transition [Fig~\ref{fig:random_saturations}.(b)].

The inverse of the dispersion of the Lanczos sequences shown in Fig.~\ref{fig:random_dispersions}
can also be divided into two groups: one shows a strong resemblance to the spectral quantum measure $\eta$ in describing the integrable-chaotic transition [Fig.~\ref{fig:random_dispersions}.(a)],
while the other [Fig.~\ref{fig:random_dispersions}.(b)] does not exhibit such a straightforward correlation.
In fact, as discussed in Appendix~\ref{sec:normalization},
when the normalization is omitted, the last group exhibits higher values which may appear to imply a chaotic behavior for all values of $k$. It remains unclear whether this specific behavior can be observed in generic systems or if it originates solely  from the random nature of this model.

\section{Final Remarks}
\label{sec:discussion}

In this work we have extended previous analyses \cite{PhysRevE.107.024217,Rabinovici_2022_integrability} about
the relationship between
Lanczos coefficient dispersion, Krylov complexity saturation and chaos
to the Schrödinger evolution of states.
In order to accomplish this, we have taken into account an alternative interpretation of the Lanczos coefficient dispersion. This enables us to expand our analysis and examine situations where diagonal disorder is present within the state formalism.
We have also investigated scenarios in which the saturation value of Krylov complexity is affected by the initial state. We found that when the initial states are widely spread in the energy eigenbasis of an integrable region, there may be an artificially high saturation value.
On the contrary, if they are sufficiently localized, it tends to be suppressed.
Both scenarios, upon initial examination, have the potential to mask the intrinsic chaotic or integrable nature of the Hamiltonian.
On the other hand, the dispersion of the Lanczos coefficients seems to be less affected by the initial state.

The situation becomes more interesting when considering initial states that correspond to eigenstates within the integrable region.
In this case, our numerical studies show that the Krylov complexity is closely aligned with the system's chaoticity measured by $\eta$.
Notably, in such cases, the normalized complexity follows almost exactly the spectral measure of chaoticity for a  significant range of parameters adjacent to those linked with the initial condition [see Figs. \ref{fig:Ising_saturations}.(a) and \ref{fig:random_saturations}.(a)].
It should be noted that local purity can also serve as a dynamic indicator of quantum chaotic behavior, although its computation is not as straightforward \cite{mirkin2021quantum}. Therefore, in our case the choice of these initial states can be seen as analogous to determining the part of the Hamiltonian with lower complexity. In other words, the complexity is evaluated in relation to a particular Hamiltonian.  In the case of the Ising model, where the region of chaotic behavior lies between two regions of integrability, even an integrable Hamiltonian can exhibit increased complexity for these initial conditions. In this scenario, the complexity arises not from the chaotic nature of the Hamiltonian itself. Consequently, it would be interesting to investigate if there exists an initial condition that can accurately track and reflect the spectral measure associated with chaos across all the range of parameters. We believe that this would provide additional insights into the intricate connection between Krylov complexity and quantum chaos.

\section{Acknowledgments}
This work has been partially supported by  CONICET (Grant No.~PIP 11220200100568CO), UBACyT (Grant  No.~ 20020170100234BA)  and ANPCyT (PICT-2020-SERIEA-01082, PICT-2021-I-A-00654,
PICT-2021-I-A-01288).

\appendix

\section{Normalization for the Lanczos coefficient dispersion}
\label{sec:normalization}

In this work we compare the quantities of interest with the chaotic measure $\eta$,
which is normalized to fit between the values 0 and 1 for an integrable and chaotic system, respectively.
In order to make these comparisons more straightforward
we would like to also map the saturation value of the Krylov complexity
and the dispersion of Lanczos coefficients to the range $[0, 1]$,
in a way that they share the same systematic as $\eta$ does with chaos.
The saturation value of the Krylov complexity
possesses a natural normalization,
since we expect $\mCK \sim 0$ for an integrable system
and $\mCK \sim \frac{D}{2}$ for a chaotic one,
but it is not the case for the dispersion of the Lanczos coefficients.

We address this issue by employing the normalization explained in Ref.~\cite{PhysRevE.107.024217}.
Given a sequence of values $X$,
we map it to that of $\eta$ by rescaling as
$X' = X \frac{\max{\eta} - \min{\eta}}{\max{X} - \min{X}}$
and applying a displacement such that fluctuations in the estimation of $\min{X}$ don't propagate to the whole rescaled sequence.
This displacement minimizes the Euclidean distance of $X'$ to $\eta$:
$d_{\min}(X', \eta) = \min_{\alpha} \norm{(X' - \alpha) - \eta}$.
Thus, the normalized sequence is defined as
\begin{equation}
    X_{\text{norm}} = X' - d_{\min}(X', \eta)
    \label{eq:normalization}.
\end{equation}

\begin{figure}[ht]
    \centering
    \includegraphics[width=\linewidth]{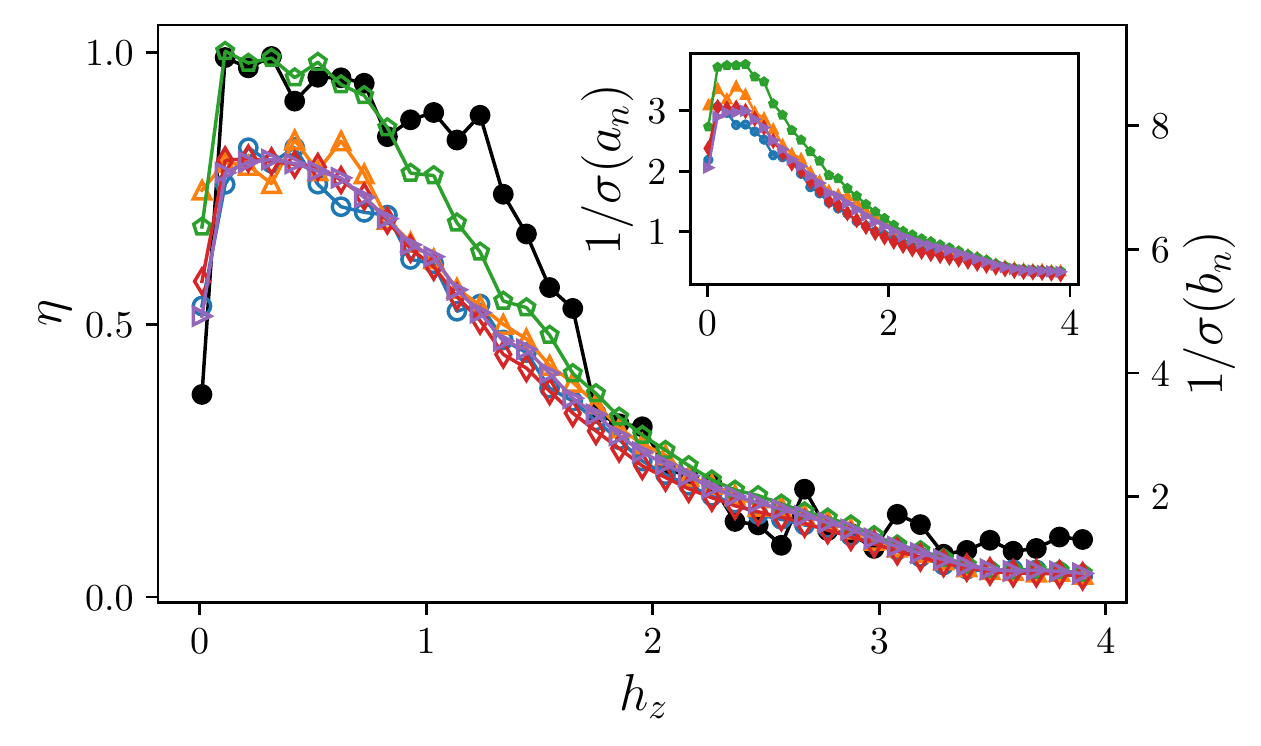}
    \caption{Chaotic measure $\eta$ (black filled circles) and (not normalized) inverse of the
        dispersion of Lanczos coefficients as a function of the chaos parameter $h_z$
        through the chaos-to-integrability transition in the Ising spin chain for a variety of initial states.
        The main plot shows the dispersion of the off-diagonal sequence $b_n$,
        while the inset that of the diagonal $a_n$
        (which has almost identical curves, save for a rescaling factor).
        The indicators for each curve are the same as in Fig.~\ref{fig:Ising_saturations}.
        }
    \label{fig:Ising_dispersions_nonorm}
\end{figure}

\begin{figure}[ht]
    \centering
    \includegraphics[width=\linewidth]{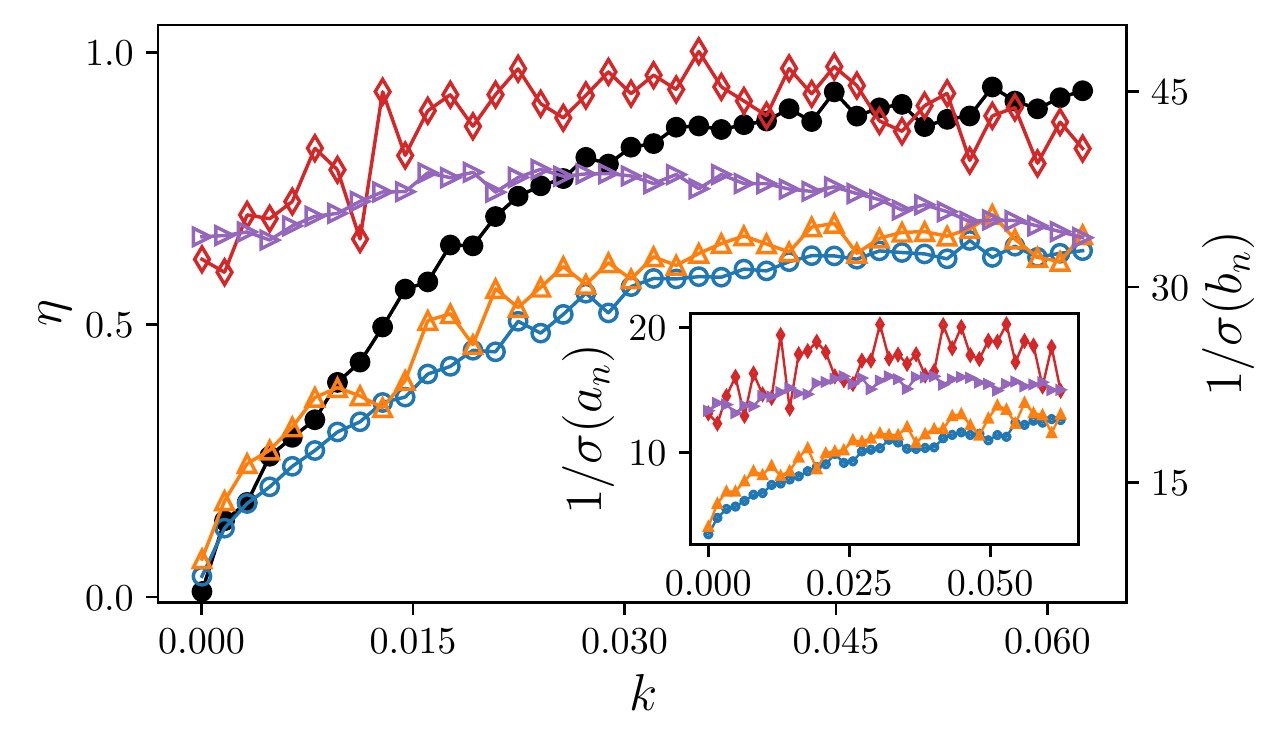}
    \caption{Chaotic measure $\eta$ (black filled circles) and (not normalized) inverse of the
        dispersion of Lanczos coefficients
        as a function of the chaos parameter $k$ through the Poisson-to-GOE ensemble transition
        for the banded random matrix model
        for a variety of initial states.
        The main plot shows the dispersion of the off-diagonal sequence $b_n$,
        while the inset that of the diagonal $a_n$
        (which has almost identical curves, save for a rescaling factor).
        The indicators for each curve are the same as in Fig.~\ref{fig:random_saturations}.
}
    \label{fig:Random_dispersions_nonorm}
\end{figure}

In the main text
[Figs.~\ref{fig:Ising_dispersions} and \ref{fig:random_dispersions}]
we show the inverse of the dispersion of the Lanczos sequences
under this normalization, i.e., $X = 1/\sigma(a_n)$ and $X = 1/\sigma(b_n)$.
Here, we show in
Figs.~\ref{fig:Ising_dispersions_nonorm} and \ref{fig:Random_dispersions_nonorm}
the same curves without normalizing,
and the conclusions of this work remain the same.
However,
since the normalization is applied separately to each curve,
this normalized dispersion measure is relative.
As a result,
while it could already be seen for the RMT model in Fig.~\ref{fig:random_dispersions}.(b)
that the curves corresponding to the uniform and random states
do not accurately follow chaoticity,
here it is made explicit that these curves show a high value of inverse dispersion irrespective of the chaos parameter $k$.


\section{Krylov complexity saturation dependence on initial state energy localization}
\label{sec:saturation_state_dependence}

In the main part of the paper we have shown numerically that both the evolution of the Krylov complexity and its saturation value are tied to the initial state of the system.
To reinforce the idea that choosing an appropriate initial state is essential to measure the complexity of a system through the Krylov approach,
we show in this appendix two limiting cases of initial states such that the saturation of Krylov complexity is almost independent on the spectrum of the Hamiltonian, and are therefore not reliable  for probing the chaoticity of the system.

The first one is an initial state completely delocalized in the energy eigenbasis,
such that
 $\abs{\braket{e_i}{\psi_0}}^2 = 1/D$ $\forall i$,
 and so Eq.~\eqref{eq:k_complexity_saturation} yields
\begin{equation}
    \mCK = \frac{D - 1}{2}
    \label{eq:k_complexity_saturation_delocalized_energy}.
\end{equation}
This result is analogous to \eqref{eq:k_complexity_saturation_delocalized},
but in that case the delocalization is of the Krylov basis elements over the energy eigenbasis.
Under operator formalism,
it was already shown in \cite{Rabinovici_2022_integrability} that
for an operator proportional to a matrix of ones in the energy eigenbasis,
the saturation of Krylov complexity is $\mCK \sim \frac{K}{2} \sim \frac{D^2}{2}$
independently of the spectrum of the Hamiltonian.
Here $D^2$ is the dimension of the operator Hilbert space.

The opposite (and trivial) scenario
is an initial state completely localized in the energy eigenbasis (an eigenstate of the Hamiltonian),
and as such a stationary state.
The Lanczos algorithm \eqref{eq:lanczos_alg} will halt immediately in this case,
yielding $b_n = 0$ and $\ket*{K_n} = 0$ $\forall n \geq 1$,
and so $C_{\mathcal{K}}(t) = 0$.
In this sense,
it may not be obvious whether the output of Lanczos algorithm is a "continuous function" of the initial state,
meaning whether or not
any initial state sufficiently localized in the energy eigenbasis
will exhibit a low (suppressed) \Kcomplexity.

To explore this, consider such a state of the form
\begin{equation}
    \psiz = \frac{1}{\sqrt{1 + \delta^2}}(\ket*{e_j} + \delta \ket*{\tilde e_j} ),
    \label{eq:psiz_delta}
\end{equation}
where $H\ket*{e_j} = e_j \ket*{e_j}$ and $\bra{e_j}\ket*{\tilde e_j} = 0$,
and the parameter $\delta$ controls the degree of localization.
Under the assumption that either
(i) the state $\ket*{\tilde e_j}$ is delocalized in the remaining orthogonal part of the energy eigenbasis
or that
(ii) the Hamiltonian $H$ can be modeled by a random matrix,
by plugging the initial state \eqref{eq:psiz_delta} into \eqref{eq:q0n}
we get, up to second order in $\delta$,
\begin{equation}
    Q_{0n} =
        \left( f_n + \frac{1}{D-1} \right) \delta^2 + \mO(\delta^4)
    \label{eq:q0n_loc}.
\end{equation}
This is so because, under the above assumption, we may write
$\abs{\braket{e_i}{\tilde e_j}}^2 \sim 1/(D-1)$
for a generic choice of $\ket*{\tilde e_j}$,
where the denominator is $D-1$ as opposed to $D$
since $\braket{e_j}{\tilde e_j} = 0$,
and so
$ \sum_{i\neq j}^D \abs{\braket{K_n}{e_i}}^2 = 1 - \abs{\braket{K_n}{e_j}}^2$.
Additionally, to obtain \eqref{eq:q0n_loc},
we show in Appendix~\ref{app:kbasis_dep_delta}
that the projection of the eigenvector $\ket{e_j}$ onto the Krylov basis follows
$\abs{\braket{K_n}{e_j}}^2 = f_n \delta^2 + \mO(\delta^4)$ for all $n \geq 1$.

Following \eqref{eq:k_complexity_saturation},
both terms in \eqref{eq:q0n_loc} must be multiplied by $n$ and summed,
which is straightforward for the second one:
$
    \sum_{n=0}^{D-1} \frac{n}{D-1} = \frac{D}{2}.
$
On the other hand, the sum of the first term is
$
    \sum_{n=0}^{D-1} n f_n,
$
and can be bounded by taking advantage of the fact that
\begin{equation}
    \sum_{n=1}^{D-1} f_n = \frac{1}{ 1+\delta^2 },
    \quad \text{with} \quad
    f_n \geq 0,
    \label{eq:norm_fn}
\end{equation}
which follows from normalization
$ \sum_{n=0}^{D-1} \abs{\braket{K_n}{e_j}}^2 = 1$,
and noting that $\abs{\braket{K_0}{e_j}}^2 = \abs{\braket{\psi_0}{e_j}}^2 = 1/(1+\delta^2)$.
From $\eqref{eq:norm_fn}$,
 we obtain the bound
\begin{equation}
    \sum_{n=0}^{D-1} n f_n \leq \frac{D-1}{1+\delta^2} = D-1 + \mO(\delta^2).
    \label{eq:bound_sum_n_fn}
\end{equation}
Finally,
we arrive to
\begin{equation}
    \mCK \leq \left( \frac{3}{2}D -1 \right)\delta^2 + \mO(\delta^4).
    \label{eq:mCK_bound}
\end{equation}

\begin{figure}[ht]
    \centering
    \includegraphics[width=\linewidth]{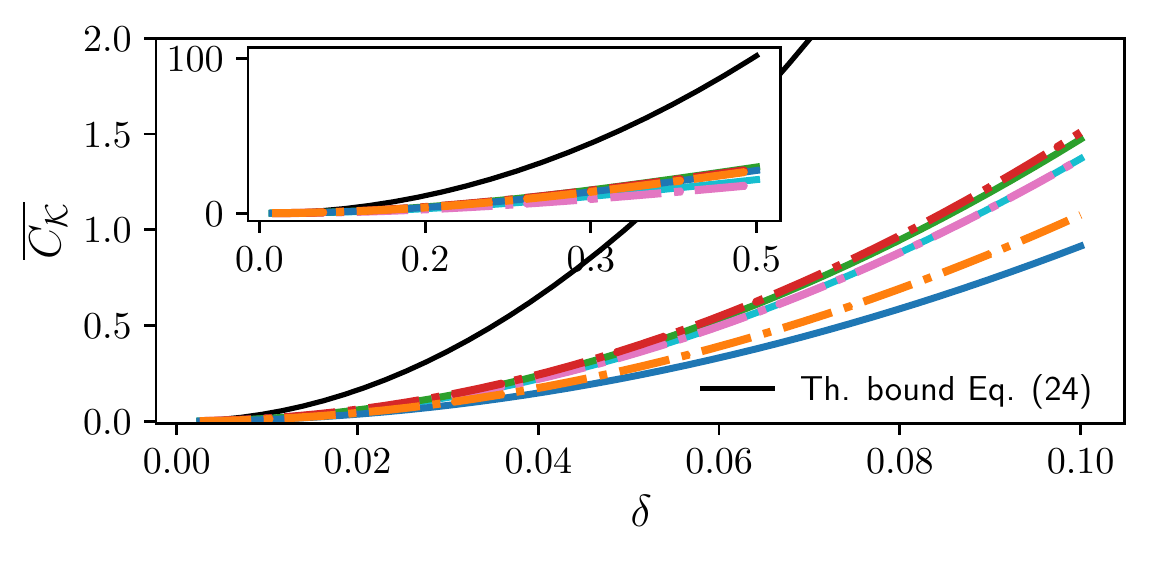}
    \caption{\Kcomplexity\ saturation for the Ising model studied in Sec.~\ref{sec:ising}.
        The inset shows the same data as the main plot, but up to larger values of $\delta$.
        The curves  were obtained for
        $j=10$, and $\ket*{\tilde e_j}$ a Gaussian profile centered at $j=61$ with standard deviation $\sigma = 10$ (blue and orange),
        $j=60$ and $\ket*{\tilde e_j}$ a Gaussian profile centered at $j=61$ with standard deviation $\sigma = 10$ (green and red),
        and
        $j=10$ and $\ket*{\tilde e_j}$ uniformly distributed across the remaining orthogonal elements of the energy eigenbasis (cyan and pink),
        where the values for the index $j$ assume an eigenbasis ordered according to increasing eigenvalues $e_j < e_{j+1}$.
        The solid lines correspond to $h_z = 4$, while the dash-dotted ones to $h_z = 0.5$.
    }
    \label{fig:Ising_lCKs_bounds}
\end{figure}

\begin{figure}[ht]
    \centering
    \includegraphics[width=\linewidth]{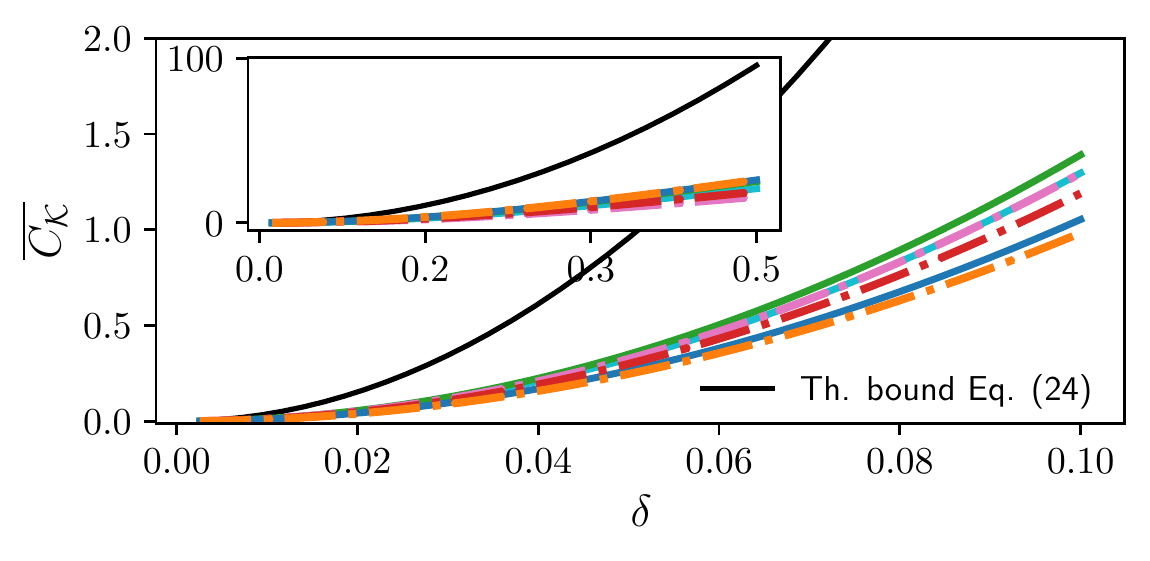}
    \caption{\Kcomplexity\ saturation for the random model studied in Sec.~\ref{sec:random}.
        The inset shows the same data as the main plot, but up to larger values of $\delta$.
        The curves were obtained for initial states
        of the same form to those considered for Fig.~\ref{fig:Ising_lCKs_bounds},
        but here the solid lines correspond to $k=0.000625$
        and the dash-dotted ones to $k=0.125$.
    }
    \label{fig:Random_lCKs_bounds}
\end{figure}

In Figs~\ref{fig:Ising_lCKs_bounds} and \ref{fig:Random_lCKs_bounds}
we show a comparison between the bound
and the saturation of Krylov complexity
for the models  considered in Secs.
\ref{sec:ising} (with $N=9$ and in the positive parity subspace)
and
\ref{sec:random} (with $D=256$),
and initial states of the form \eqref{eq:psiz_delta}.
This was done for different choices of $j$ in $\ket*{e_j}$ and $\ket*{\tilde e_j}$,
for the latter both delocalized and localized
(such that the assumptions above don't necessarily apply),
and in the integrable and chaotic regimes.
In all cases the bound holds,
up to rather large values of $\delta$.
The complexity curves get closer to the bound
when both $\ket*{e_j}$ and $\ket*{\tilde e_j}$ are chosen near the center of the energy spectrum.
The lowest saturation values are obtained when these  are far apart from the center.

These results highlight the fact that
in order to examine the overall chaoticity of the system through the use of the \Kcomplexity, the appropriate choice of initial state may not be so obvious. For instance, an arbitrary initial state will be delocalized in the energy eigenbasis, leading to an "artificially" high complexity saturation.
On the contrary, simpler states written in terms of the energy eigenbasis elements have their complexity saturation suppressed, obscuring the chaotic nature of the underlying Hamiltonian.
This characteristic behavior with the initial state has also been observed in other metrics of quantum complexity, including the Loschmidt echo \cite{gorin2006dynamics,jacquod2009decoherence,Goussev:2012} and out-of-time-ordered correlators \cite{GarciaMata2022}.


\section{Dependence of the Krylov basis on the degree of localization}
\label{app:kbasis_dep_delta}
In this appendix we show that for an initial state of the form of Eq.~\eqref{eq:psiz_delta},
the projection of the Krylov basis elements onto the eigenstate $\ket*{e_j}$
follow
$\abs{\braket{K_n}{e_j}}^2 = f_n \delta^2 + \mO(\delta^4)$ with $f_n \geq 0$ for all $n \geq 1$,
which is a necessary step in order to obtain the bound of Eq.~\eqref{eq:mCK_bound}.
We do this by explicitly carrying out the Lanczos algorithm \eqref{eq:lanczos_alg}
to the considered initial state up to $n=3$,
and then proceeding by induction.

Applying the Lanczos algorithm [Eq.~\eqref{eq:lanczos_alg}] to the initial state
Eq.~\eqref{eq:psiz_delta} up to $n=3$ we obtain, for both $n=2, 3$,
\begin{equation}
\begin{split}
        & a_n = a_n^{(0)} + a_n^{(2)}\delta^2 + \mO(\delta^4), \\
        & b_n = b_n^{(0)} + b_n^{(2)}\delta^2 + \mO(\delta^4), \\
        & c_n = c_n^{(0)} + c_n^{(2)}\delta^2 + \mO(\delta^4)
         \label{eq:app_delta_dep_1}
\end{split}
\end{equation}
and
\begin{eqnarray}
        \ket*{K_n} &=&
        (p_n^{(1)}\delta + p_n^{(3)}\delta^3) \ket*{e_j}
        + \nonumber \\
        && (q_n^{(0)}\delta + q_n^{(2)}\delta^2) \ket*{\tilde e_j}
        +
        \mO(\delta^4),
         \label{eq:app_delta_dep_2}
\end{eqnarray}
\noindent where $c_n \equiv \ev{H^2}{K_n}$,
and both $p_n^{(m)}$ and $q_n^{(m)}$ are functions of the Hamiltonian $H$.
For $n=1$ these relations hold,
except for $b_1 = b_1^{(1)}\delta + b_1^{(3)}\delta^3 + \mO(\delta^5)$.
For the purposes of this derivation,
we are not concerned about the particular values the expansion coefficients $p_n^{(m)}$ and $q_n^{(m)}$ may take.

Now we proceed by induction.
Suppose that for some $n$ the relations in Eqs.~\eqref{eq:app_delta_dep_1} and \eqref{eq:app_delta_dep_2}
hold for both $n$ and $n-1$.
By applying the Lanczos algorithm once again for the next iteration step,
it is straightforward to check that these relations also hold for $n+1$.
Thus, since they hold for $n=2$ and $3$,
they do also for $n=4$, and so, hold for all $n \geq 2$.

In particular, we have
$\ket*{K_n} =
(p_n^{(1)}\delta + p_n^{(3)}\delta^3) \ket*{e_j}
+
(q_n^{(0)}\delta + q_n^{(2)}\delta^2) \ket*{\tilde e_j}
+
\mO(\delta^4)$ $\forall n\geq 1$,
and so
\begin{equation}
    \abs{\braket{K_n}{e_j}}^2 = {p_n^{(1)}}^2 \delta^2 + \mO(\delta^4)
    \quad \forall n\geq 1
    \label{eq:app_kn_delta_dep},
\end{equation}
where we may identify $f_n = {p_n^{(1)}}^2 \geq 0 $.


\bibliography{references}
\end{document}